# Production of Phase Screens for Simulation of Atmospheric Turbulence

Rachel Rampy[1], Don Gavel[2], Daren Dillon[1], Sandrine Thomas[2]
[1]*University of California, 1156 High St., Santa Cruz, CA, USA 95064-1099*
[2]*University of California Observatories*

The ability to simulate atmospheric turbulence in the lab is a crucial part of testing and developing astronomical adaptive optics technology. We report on the development of a technique for creating phase plates, which involves the strategic application of clear acrylic paint onto a transparent substrate. Results of interferometric characterization of these plates is described and compared to Kolmogorov statistics. The range of $r_0$ (Fried's parameter) achieved thus far is 0.2 – 1.2 mm, with a Kolmogorov power law. These phase plates have been successfully used by the Lab for Adaptive Optics at University of California, Santa Cruz, in the Multi-Conjugate Adaptive Optics testbed, as part of the Villages (Visible Light Laser Guidestar Experiments) calibration system, and during integration and testing of the Gemini Planet Imager. This method has proven to be an effective and low cost means to simulate turbulence. We are now distributing the plates to other members of the AO community.

## 1. Introduction

As the field of adaptive optics (AO) continues to expand in new directions, the technological prowess needed for these endeavors is rapidly increasing. From the onset, AO was established to compensate the effects of turbulent air in Earth's atmosphere on propagating electromagnetic fields. Hence, an understanding of how turbulence manifests in astronomical images is critical. Research into the properties of turbulent media and energy exchange in the atmosphere led to the development of Kolmogorov[1] and frozen flow theory[2]. Respectively, these enable a means of assessing spatial-temporal properties of the phenomenon, and allow reasonable modeling using plates with stationary phase aberrations.

For testing and calibrating a complex adaptive optics (AO) system it is useful to have an artificial turbulence generator with known, realistic, and repeatable characteristics. Many current schemes for creating representative atmospheric aberrations are costly and place cumbersome requirements on the optical system. One key technical challenge is producing turbulence with $r_0$ (Fried's parameter[3]) values as small as a few hundred micrometers at visible wavelengths. A larger $r_0$ requires a correspondingly large pupil diameter, and larger relay optics, in order to achieve proper statistics for telescopes of modern size. Several methods of generating turbulence that have been investigated and reported in the literature include near index matching[4], hot air chambers[5], liquid filled chambers[6], spatial light modulators[7], ion-exchange phase screens[8], and surface etching[9].

In this paper we present a method of creating turbulent phase plates that was developed in the Laboratory for Adaptive Optics (LAO) at the University of California, Santa Cruz. It is capable of producing a variety of strengths and characteristics, and has proven to be a successful means of injecting turbulence into optical systems operating at infrared and visible wavelengths. The technique involves the strategic application of clear acrylic paint onto a transparent substrate, such as glass or plastic. The next section contains an outline of this technique and how it was developed, followed by a description of how the properties of the materials used and resulting plates were assessed, in section 3. Section 4 contains results from plates of varying turbulence strength, and a comparison with computer generated Kolmogorov phases. In section 5 the outcome of using these plates in an on-sky AO testbed is presented, with a comparison to real atmospheric turbulence measured with the same instrument, and a brief analysis of how often the sky follows Kolmogorov statistics. Section 6 discusses the overall effectiveness of this method, including issues of repeatability and the diversity of results.

## 2. Fabrication

The current technique is a derivative of the work of Thomas[10]. The method initially involved spraying hairspray onto glass plates, where variations in the thickness of the applied material create optical path differences similar to that of the atmosphere. In our investigation, it was soon apparent that a non water-soluble spray would be preferable to avoid absorbing moisture from the environment and collecting dust. Experimentation began with Krylon brand clear acrylic paint. After the fabrication of several phase plates by hand we designed an automated system to help with controlling variables involved in the technique.

The automated method involves moving the spray can on an x-y stage over the plate, using two Velmex Inc. slides mounted upside-down. Attached to the lower slide is a harness that holds the aerosol can and a solenoid to turn it on and off. The assembly is enclosed in a plexiglass box with a ventilation system, consisting of two air filters and a fan, to help circulate air and remove particulates and paint fumes. The entire setup is housed in a fume hood for further elimination of hazardous paint fumes from the work environment. Figure 1 contains a diagram of the apparatus, and Figure 2 shows photos of the interior and exterior.

Layers of paint are applied to a stationary plate in a raster scan pattern, as shown on the right in Figure 1. The machine allows precise control of variables such as how quickly the can moves, distance between back and forth passes, and how far the can is above the plate. Other factors that affect the statistics of the imprinted turbulence include the number of layers and how much time elapses between applications. The level of airflow around the plate while the paint is drying has also shown to impact the final product. Too much wind, such as from leaving the fan on in the box, can result in bubbles being trapped in the paint or the formation of directional high order structure. By taking advantage of the full dynamic range of the linear motors, we have successfully used this machine to produce phase plates with diameters up to 20 cm. The standard size substrates we offer are 7.6 cm (3 inch) and 14 cm (5.5 inch) diameter disks.

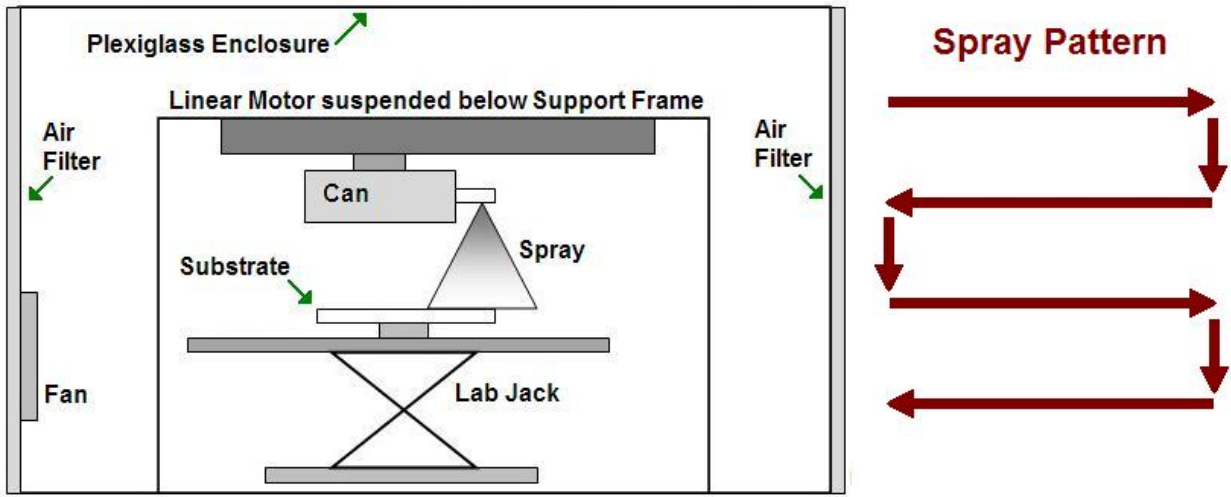

**Figure 1:** A diagram of the spraying apparatus is shown on the left, and the arrows on the right indicate the path followed by the can while coating the substrate.

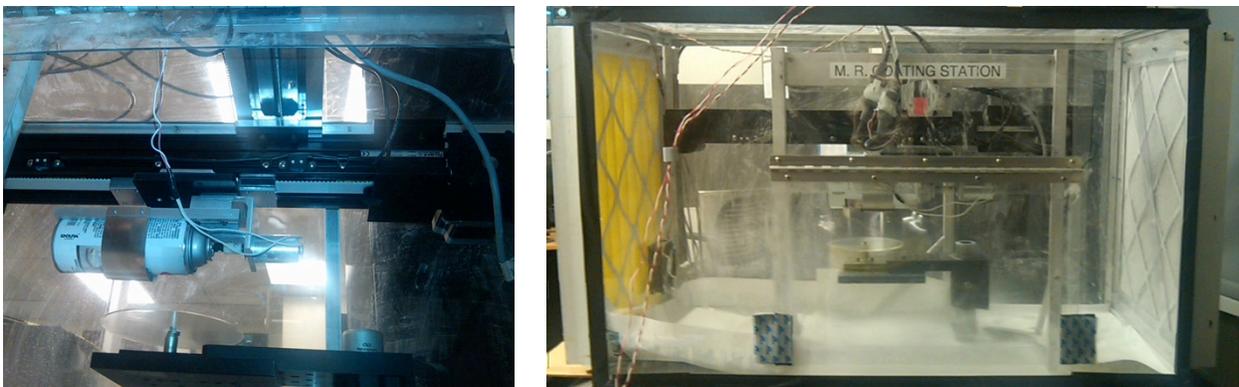

**Figure 2:** The left photograph shows the aerosol can suspended from the two motorized slides. The plate being sprayed is also visible beneath the can, near the bottom left. On the right is the entire setup, as viewed from outside the plexiglass enclosure.

In most AO testing applications using phase plates we want to model the effects of dynamic turbulence, i.e. wind. This can be achieved by translating a rectangular plate through the test beam path, or by rotating a circular disk for continuous turbulence. For a lightweight and durable disk we chose to use 3 mm thick, $\lambda/10$ optical quality, plastic plates manufactured by Edmund Optics. These substrates are made of Thermoset ADC (CR-39®), where the thermosetting process generates very little birefringence; the plastic has an Abbe Value of 59, which is close to that of glass (which is 60) and has an index of refraction equal to 1.5[11]. Acrylic, used in the spray, has a refractive index of 1.49 at 588nm and an Abbe number of 55.3. These characteristics enable the phase plates, while not perfectly achromatic, to have properties close to that of a clear glass such as BK7. This material is also receptive to the process of laser cutting, which allows the flexibility to create turbulence plates whose dimensions have been tailored to meet the specific needs of the user.

# 3. Analysis

Here we present the process used in selecting optimal materials for this type of turbulence plate, and describe how the resulting performance is assessed. There are several varieties of transparent spray paint, so it was necessary to investigate the potential of each for this application. The chromatic transmission of acrylic paint and the plastic and glass substrates is examined. The section concludes with a description of measuring the phase aberrations in a way that allows high-resolution spatial mapping.

Seven varieties of clear spray paints where investigated. Table 1 lists the type of spray, the amount of scintillation it produced on a heavily coated glass plate, along with various other pros and cons. Intensity variations were measured by a relay consisting of two lenses and a Basler CCD camera. The plate scale with this arrangement was 162 pixels per millimeter. The scintillation with nothing in the system was 2%, and the uncoated plastic and glass substrates were both found to have 5% variation in intensity.

| Type of Spray | % Intensity Variation | Pros | Cons |
|---|---|---|---|
| **Krylon Acrylic** | 9 | Completely transparent | Low viscosity, difficult too apply an even coating |
| **Krylon Acrylic UV** | 11 | | |
| **8321 Lacquer** | 14 | None | Yellow tint |
| **Minwax Lacquer** | 10 | | |
| **Ace Polyeurothane** | 8 | None | Yellow tint |
| **Krylon Polyeurothane** | 10 | | |
| **Ace Enamel** | 48 | Higher viscosity, easy to apply an even coating | Contains pigment which causes large intensity variations |

**Table 1:** A comparison of different aerosols showed the Krylon Acrylic spray to be the optimal choice because it is completely transparent, and does not cause significant intensity variations.

The lacquer and polyeurothane had similar viscosity to the acrylic sprays, but showed a slight yellow tint after several layers were applied. The enamel spray was very easy to work with because of its higher viscosity. However, multiple layers showed a slight increase in opacity. The scintillation measurements, and images taken with a Zygo interferometer, confirmed the presence of light scattering pigments in the spray. Figure 3 shows these results from the Zygo, where the left image is of an uncoated plastic disk and the right is a disk heavily coated with the enamel. The irregularities visible in the right image indicate the presence of light scattering particles.

The acrylic is the optimal choice since it is completely transparent, even after the application of several thick layers, and induces only about 10% scintillation, which is less than seen in the normal atmosphere.

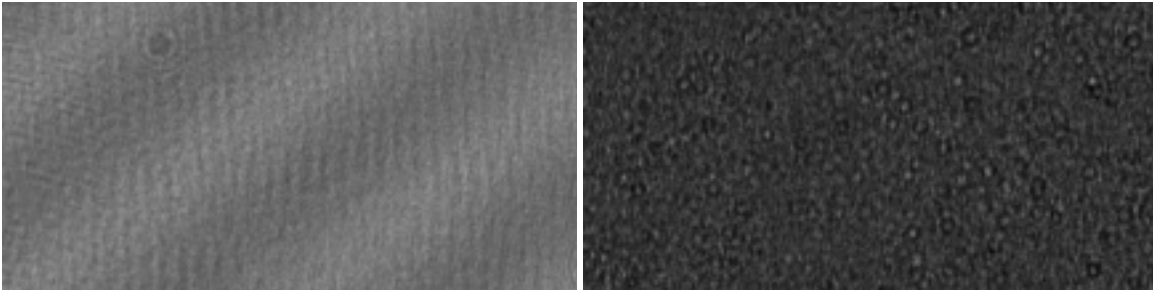

**Figure 3:** Images from the Zygo interferometer show the presence of light scattering pigments in the enamel spray. The left image is an uncoated plastic disk, and on the right is a disk coated with the Ace enamel, at the same magnification. The "ripples" seen on the uncoated disk are interference patterns due to the disk surface not being perfectly perpendicular to the beam.

The transmission of the substrates and acrylic spray were measured between 200 and 2000 nm as shown in Figure 4. The plastic has a significant decrease around certain near infrared wavelengths. It is not expected that this will interfere with overall functionality on a testbed or as a simulator in an on-sky AO system, since the light source intensity can generally be adjusted in these situations and testing is usually done at wavelengths shorter than 1600 nm.

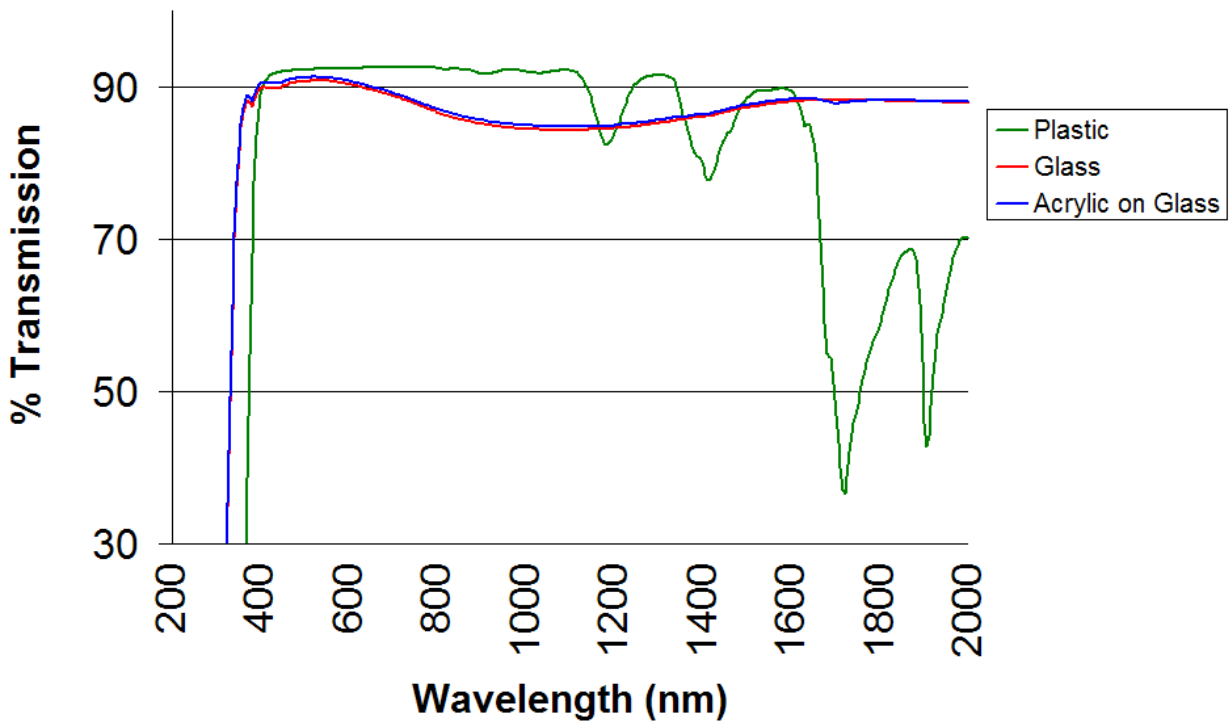

**Figure 4:** Transmission spectra for the two substrates, plastic (green) and glass (red), and for the acrylic spray on glass (blue). The presence of the spray on glass causes no significant decrease in transmission, and the slight increase (blue line is slightly above the red) is likely due to lensing by the uneven surface of the acrylic.

Measurements of the wavefront distortion were preformed with a Quadrature Polarization Interferometer (QPI), which operates with a 632 nm HeNe laser. This is a Mach-Zehnder interferometer that measures the phase of an optic in single pass transmission[12]. The cameras are setup so the scale is 67 pixels per millimeter at the location of the plate, giving a resolution of about 15 μm. Maps of the phase were analyzed by taking the spatial power spectrum and structure function and comparing them to predictions of Kolmogorov theory[1]. This is discussed in more detail in the next section.

## 4. Results

The objective of this work is to produce turbulence plates with values of $r_0$ in the range 300 – 800 μm, measured at a wavelength of 632 nm (or equivalently, about 225 – 605 μm measured at a wavelength of 500 nm). This was set by our need to simulate 8 – 10 m class telescopes using test beam sizes of only a few centimeters. Through trial and error, methods to realize these parameters uniformly across a plate have been developed. We present four plates, two exemplifying this range of $r_0$, and two that show the minimum and maximum $r_0$'s created thus far, which are 200 and 1200 μm measured with 632 nm light. These are compared to phase maps generated by an atmosphere simulation code[13].

The plots in this section (Figures 5 through 9) have the predictions of Kolmogorov theory shown as red dashed lines, and the phase data are thin black lines. The phase structure function, $D_\Phi$, is given by

$$D_\Phi(r) = <[\Phi(x) - \Phi(x+r)]^2>,$$

where $\Phi(x)$ is the value of the phase at location $x$, and $\Phi(x+r)$ is the value of the phase at a distance $r$ from location $x$. The structure function is just the difference between the phase at two locations separated by a distance $r$, squared and averaged over the ensemble. For Kolmogorov turbulence this is

$$D_\Phi(r) = 6.88 \left(\frac{r}{r_0}\right)^{5/3}.$$

The power spectrum for Kolmogorov turbulence is

$$P(k) = 0.027 r_0^{-5/3} k^{-11/3},$$

where $k$ represents spatial frequency. The analysis is done by squaring the Fourier Transform of the phase map and taking a radial average. On a log-log scale the graph of the power versus frequency is expected to have a slope of -11/3, and the structure function versus the separation, r, has a 5/3 slope. Where the data roll off near the edges of the graphs is an indication of the outer scale imprinted on the plate, which vary from 1 – 4 mm. In the power spectrum, a larger $r_0$ would correspond to shifting the dotted line downwards, while in the structure function the opposite is true.

Plates A and B (Figures 5 and 6) exhibit how Kolmogorov-like turbulence profiles have been achieved within the desired range. They are 14 cm and 7.6 cm diameter disks, respectively, and each had 8 locations tested with QPI. The regions analyzed are circular with 1.4 cm diameter. Three example images of reconstructed phase are shown for each plate below the graphs. These are useful in verifying the uniformity of the turbulence at different locations on the plate.

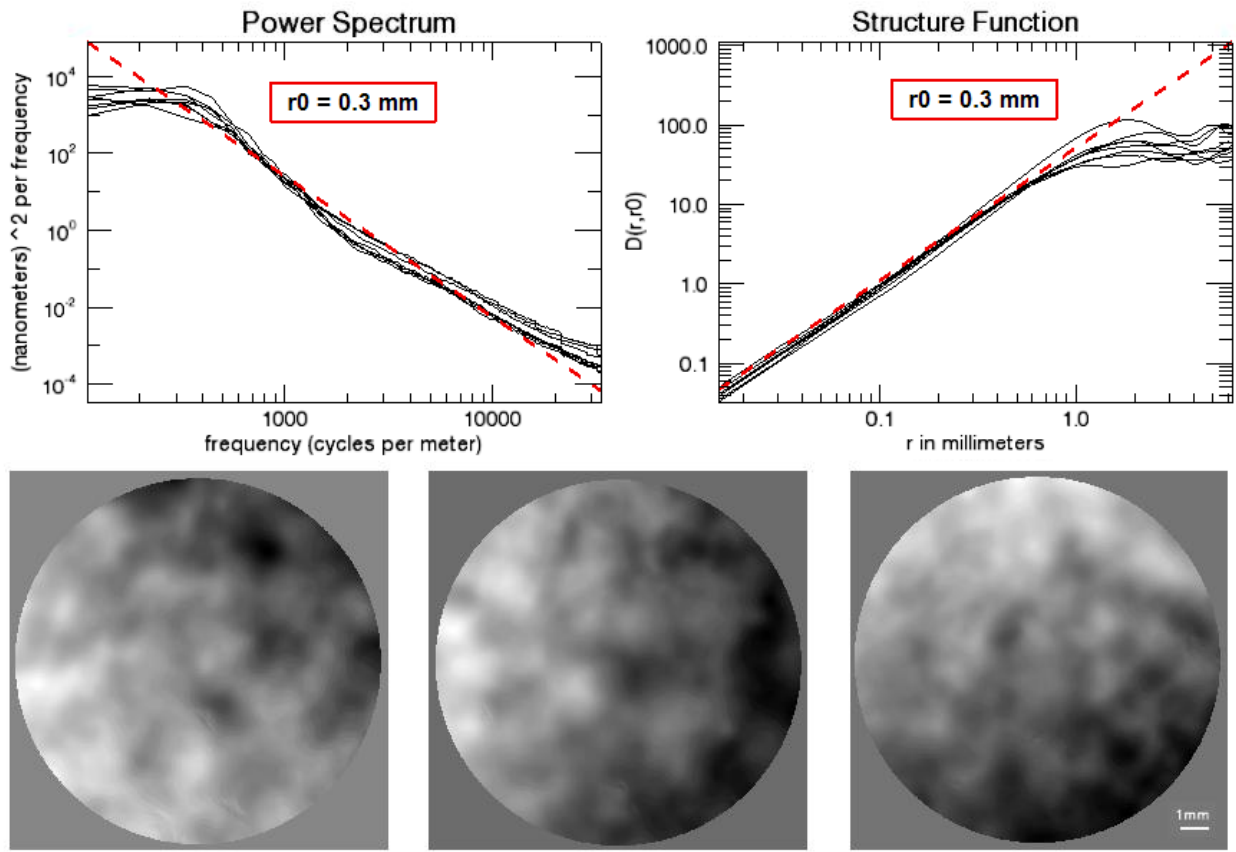

**Figure 5:** This shows the resulting structure function and radially averaged power spectrum for 8 locations on a 14 cm diameter disk. The dashed red lines represents Kolmogorov theory for $r_0 = 0.3$ mm, at 632 nm wavelength. The bottom images show reconstructed phase from the Quadrature Polarization Interferometer (QPI) for 3 of the data points. These are each 1.4 cm across, and verify the uniformity of the turbulence across the plate.

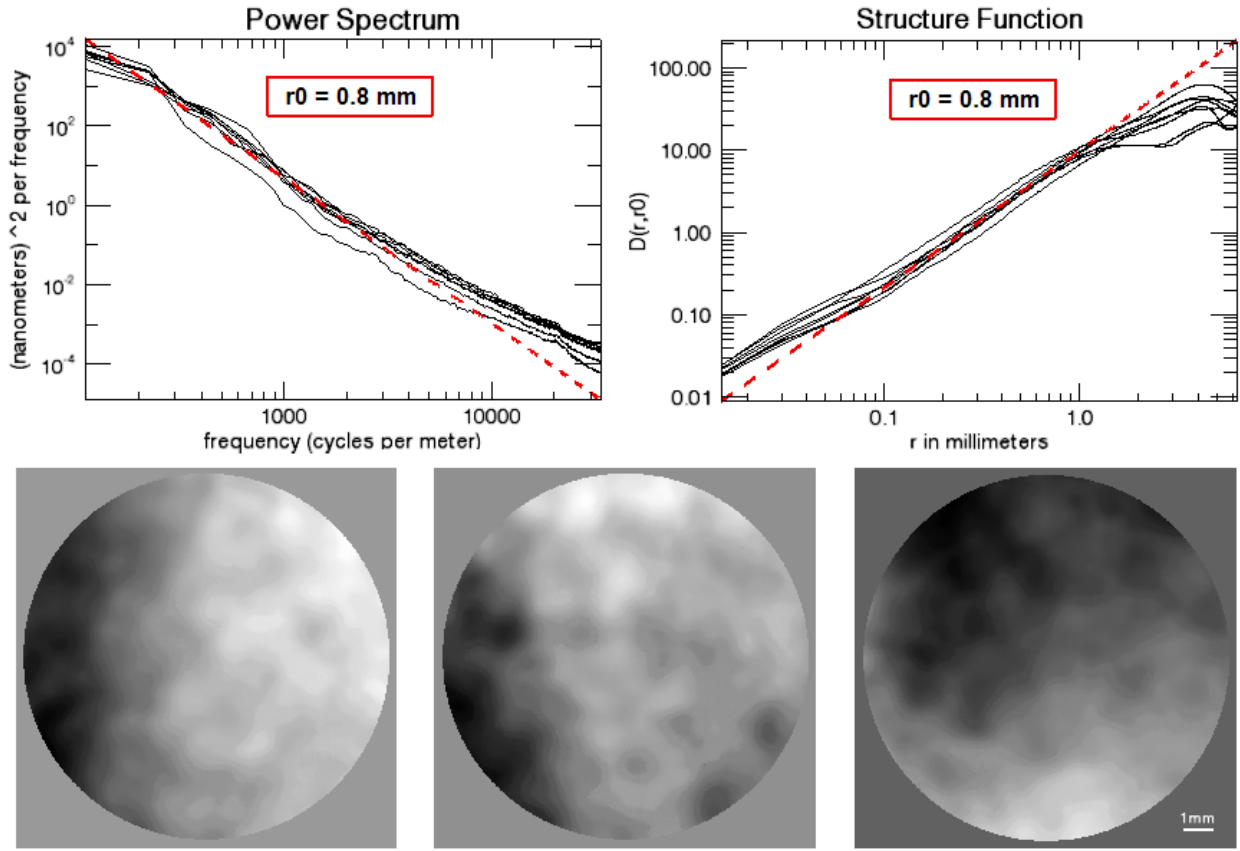

**Figure 6:** This shows the resulting structure function and radially averaged power spectrum for 8 locations on a 7.6 cm diameter disk, where the red lines represents Kolmogorov theory for $r_0 = 0.8$ mm, at 632 nm wavelength. The bottom images show reconstructed phase from QPI for 3 of the locations tested. These are an indicator of the uniformity of the turbulence.

For comparison, and to ensure validity and proper calibration of the power spectrum and structure function programs, an identical analysis was carried out on computer generated phase screens. Figure 7 shows the structure function and power spectrum for eight versions of phase created from different "seed" values in the simulation program, and 3 phase images. The different iterations show some variability, as is also seen in the handmade plates, and fit the theoretical slopes and target $r_0$ very well.

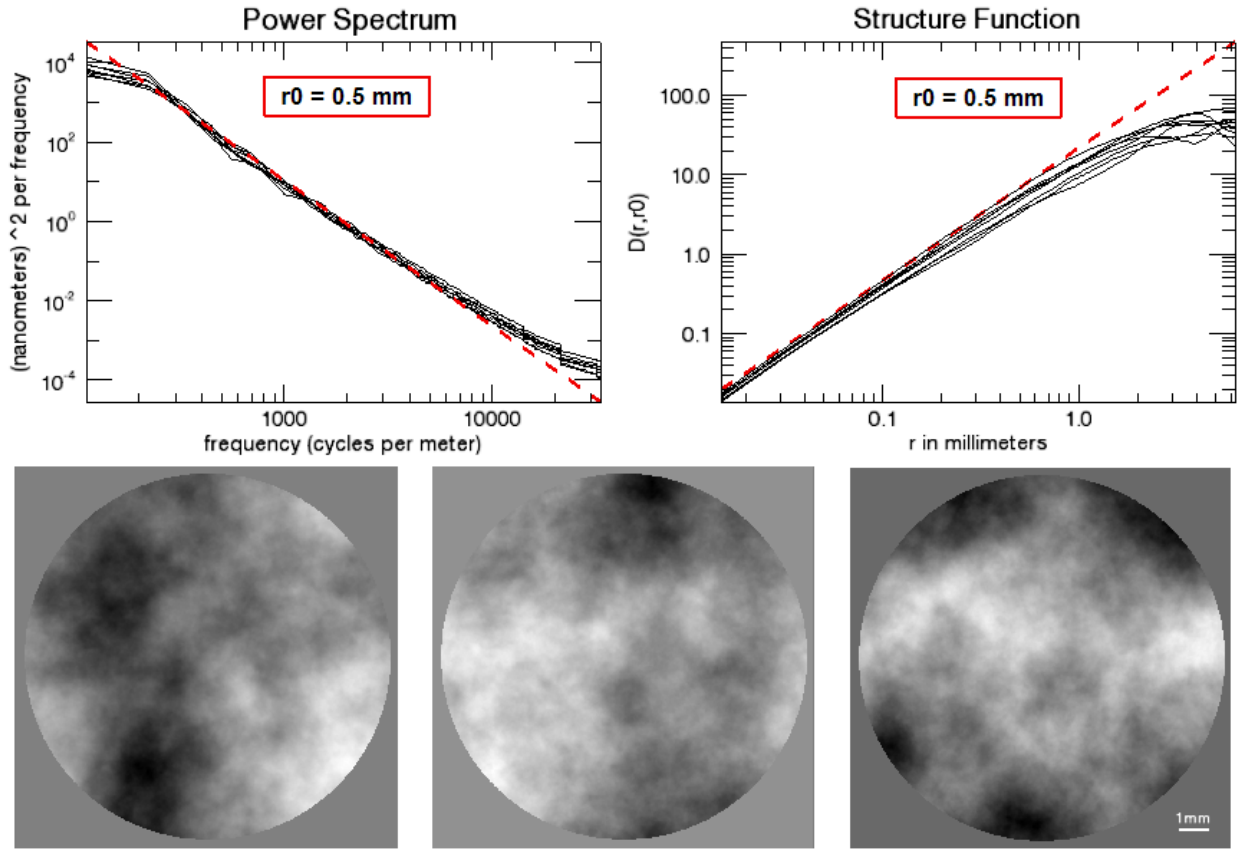

**Figure 7:** Eight versions of computer generated phases were evaluated, three of which are shown. The similarity between these and the physical plates indicates our analysis is valid.

The effort of extending our method to creating plates with smaller and larger values of $r_0$ has met some success, although generally these plates have less well defined statistics and are more prone to cosmetic issues, like unevenness. It has also been much more difficult to produce these more extreme values repeatedly. Plates C and D (Figures 8 and 9) are examples of the minimum and maximum $r_0$'s we've achieved. These follow Kolmogorov statistics somewhat less well, and exhibit greater variability from one location to another. This method has also been somewhat successful at producing low order only and high order only turbulence profiles. These are useful for testing a woofer-tweeter AO system configuration, such as that used in Gemini Planet Imager AO system[14].

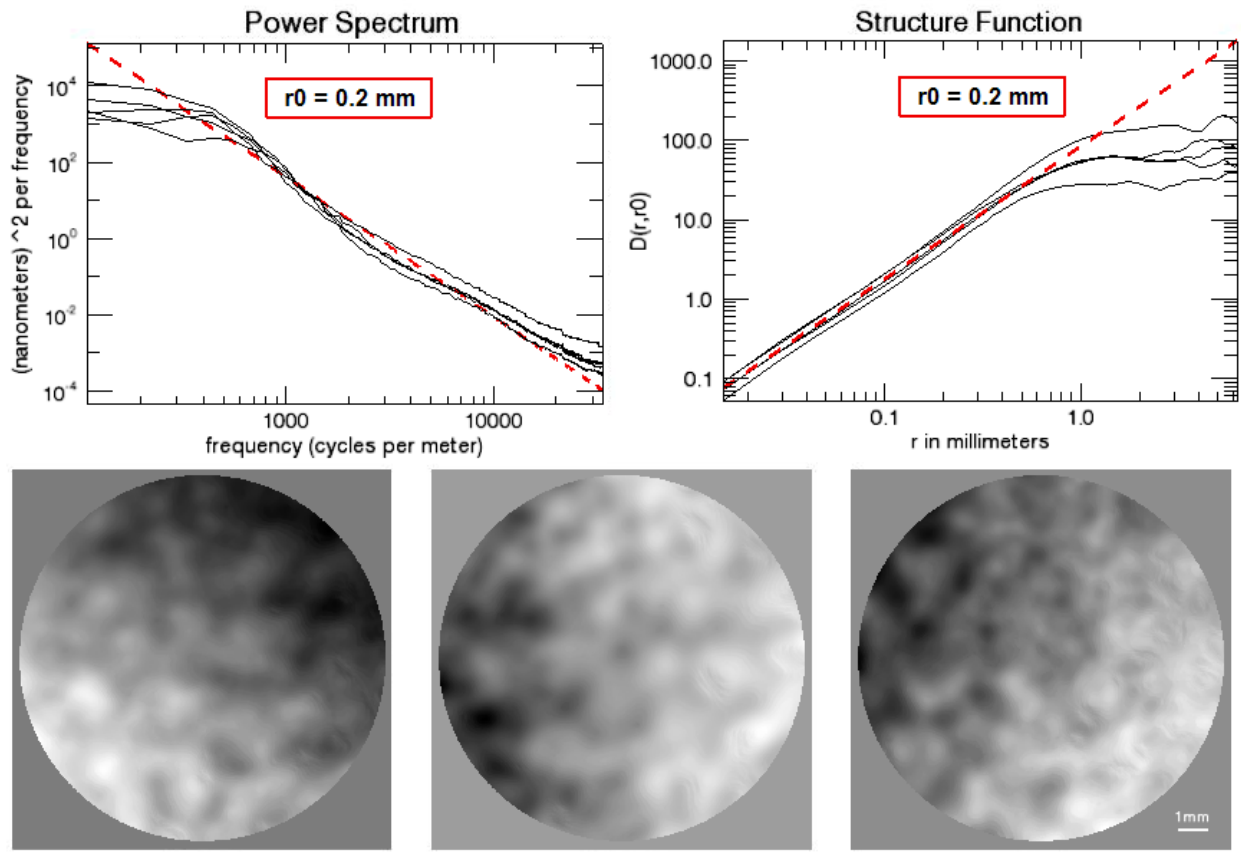

**Figure 8:** This shows the resulting structure function and radially averaged power spectrum for 5 locations on a 14 cm diameter disk. The dashed red lines represents Kolmogorov theory for $r_0 = 0.2$ mm, at 632 nm wavelength. The bottom images show reconstructed phase from the Quadrature Polarization Interferometer (QPI) for 3 of the data points.

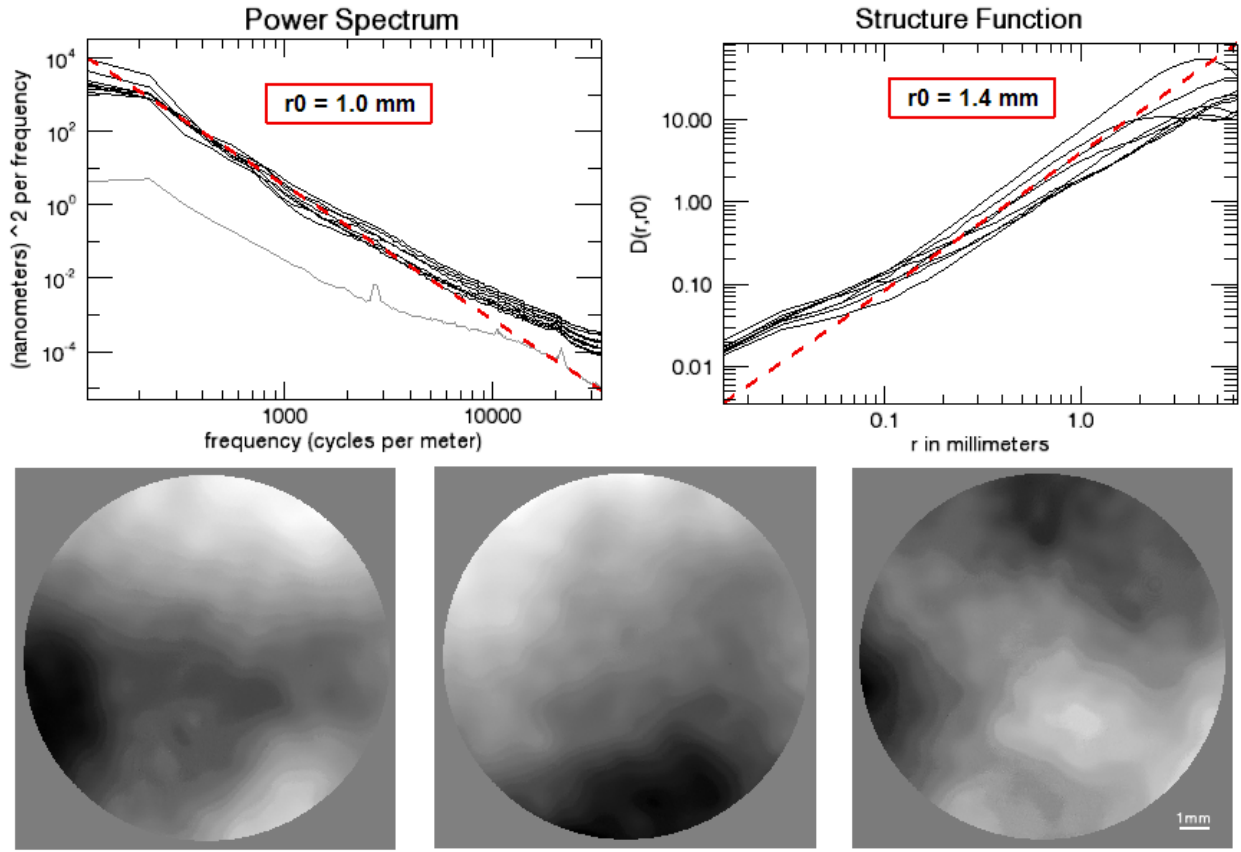

**Figure 9:** This shows the resulting structure function and radially averaged power spectrum for 8 locations on a 14 cm diameter disk. The dashed red lines represents Kolmogorov theory for $r_0 = 0.8$ mm, at 632 nm wavelength. The bottom images show reconstructed phase from the Quadrature Polarization Interferometer (QPI) for 3 of the data points. The gray line towards the bottom of the power spectrum represents the noise present in QPI. A remnant of the vibration near $2 \times 10^4$ cycles/m is evident in the data here, and also in plate C.

It is interesting to note that the larger $r_0$ plates tend to show greater excess power at the high frequency end of the spectrum. A slight surplus is expected as a remnant of taking the Fourier Transform over a finite aperture, and is seen in both real and computer generated phase data. The additional excess in the large $r_0$ phases is likely due to noise in the QPI measurements. A measurement of this noise is shown against the power spectrum of plate D in Figure 9. Residuals of the vibration near $2 \times 10^4$ cycles/m are seen in the data here and in plate B (Figure 6).

In table 2 the values $r_0$, RMS, and Peak-to-Valley, of the four plates and simulations are tabulated with one standard deviation of the samples, all measured at a wavelength of 632 nm. These numbers show the expected trends of increased RMS and Peak-to-Valley for plates with smaller $r_0$'s, i.e. stronger turbulence. The values for the 10 iterations of computer generated phases show good agreement with the manufactured plates.

|  | $r_0$ (µm) | RMS (µm) | Peak-to-Valley (µm) |
| --- | --- | --- | --- |
| **Plate A** | 0.3 | 0.54 ± 0.09 | 3.42 ± 0.67 |
| **Plate B** | 0.8 | 0.44 ± 0.07 | 2.59 ± 0.50 |
| **Simulations** | 0.5 | 0.48 ± 0.08 | 2.81 ± 0.44 |
| **Plate C** | 0.2 | 0.72 ± 0.17 | 4.67 ± 1.16 |
| **Plate D** | ~1.2 | 0.30 ± 0.10 | 1.74 ± 0.45 |

**Table 2:** The values of $r_0$, RMS, and Peak-to-Valley determine the strength of the turbulence on a given plate. The RMS and Peak-to-Valley are given with ± one standard deviation. The computer generated phases (Simulations) agree well with what is found for the manufactured plates.

## 5. Performance in an on-sky AO testbed

The Visible Light Laser Guidestar Experiments (Villages) is a Micro-Electro Mechanical Systems (MEMS) based visible-wavelength adaptive optics testbed on the Nickel 1-meter telescope at Lick Observatory. Its purpose is to compare the effectiveness of open and close loop control methods with a MEMS device[15]. The unique layout of this instrument allows simultaneous evaluation of corrected and uncorrected (not incident on the deformable mirror) paths. It is also capable of switching to an internal fiber light source and turbulence plate for testing and calibration purposes. This feature was used to compare the temporal power spectrum of one of our plates with actual sky turbulence. In addition, a brief analysis to determine how well the sky follows Kolmogorov statistics was carried out.

Data taken with Villages between June and November of 2010 were inspected. Figure 10 shows a temporal power spectrum of the phase plate, the general shape of which was constant in time and fits the desired statistics well. To analyze the sky's power spectrum, 40 sets of data from 7 different nights were examined by eye. It was found that they fall into three categories, either shallow (i.e. a lack of power at low frequencies and an excess at the high end), bowed out (an excess of power at mid frequencies), or followed Kolmogorov theory well. Examples of each of these are shown in Figures 11 – 13, respectively. Each type occurred in approximately equal numbers, meaning the sky was only Kolmogorov in a third of the data sets.

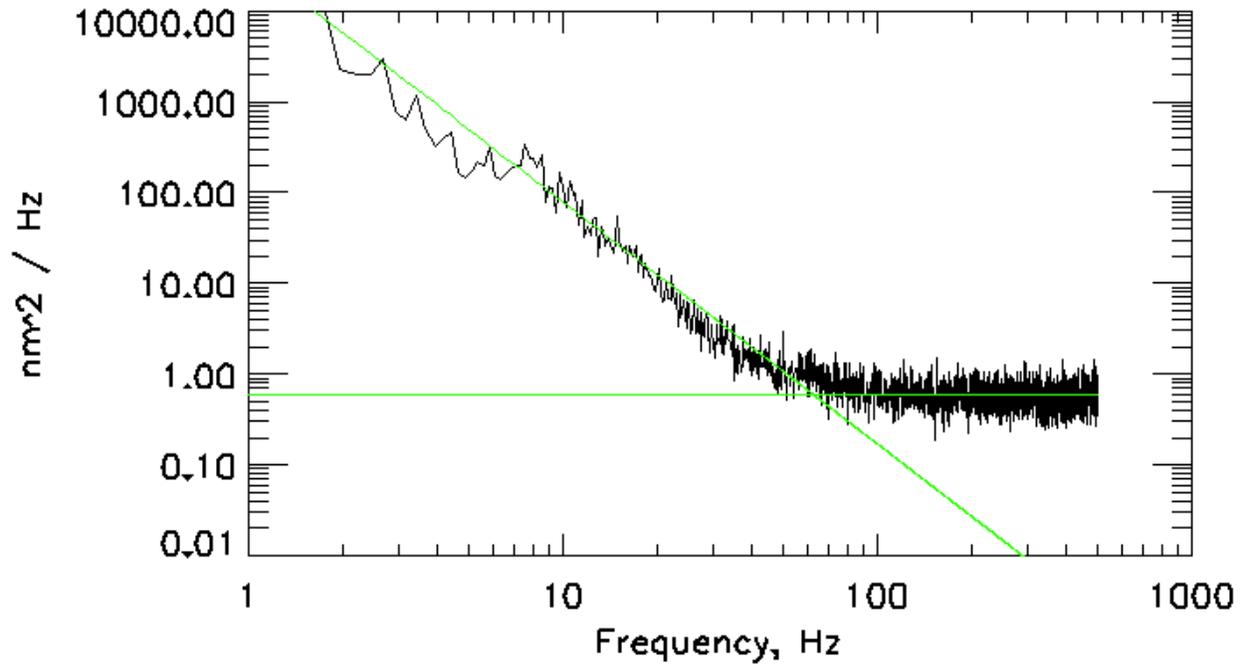

**Figure 10:** Power spectrum of the turbulence plate in the Villages instrument. The black line is measurements from the open loop path, the horizontal line is the noise floor and the tilted line represents Kolmogorov theory, which the data matches very well.

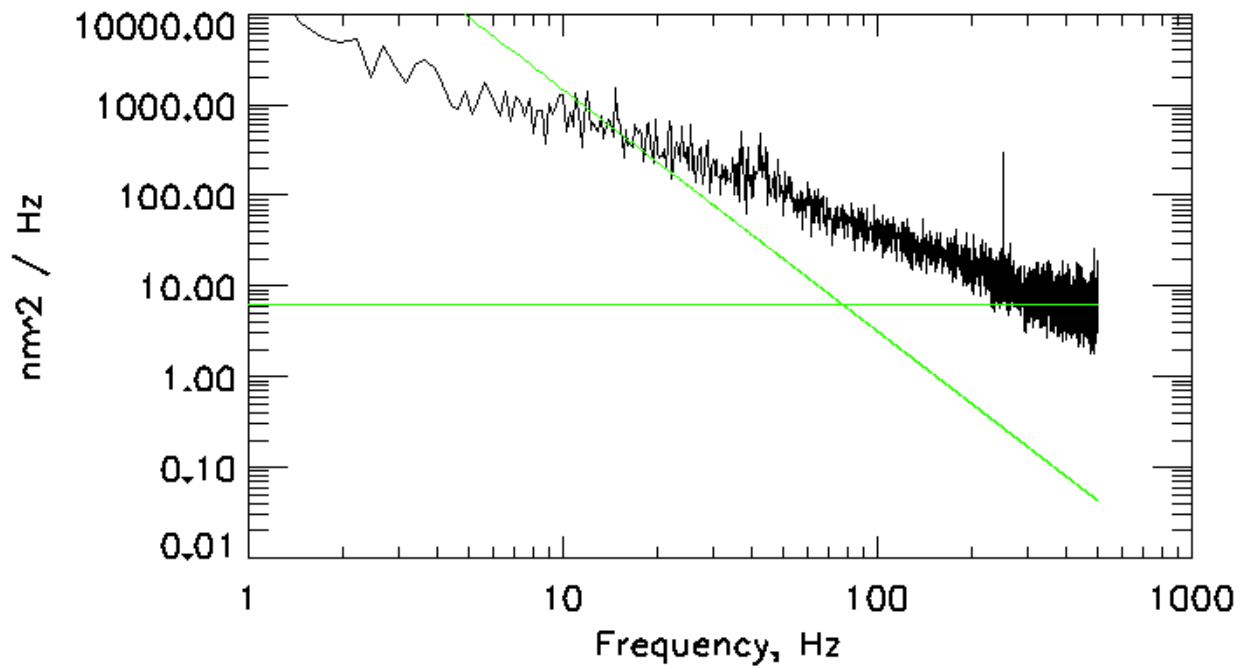

**Figure 11:** Example of shallow sky power spectrum. The black line is measurements from the open loop path, the horizontal line is the noise floor and the tilted line represents Kolmogorov theory. This occurred in about a third of the 40 data sets examined.

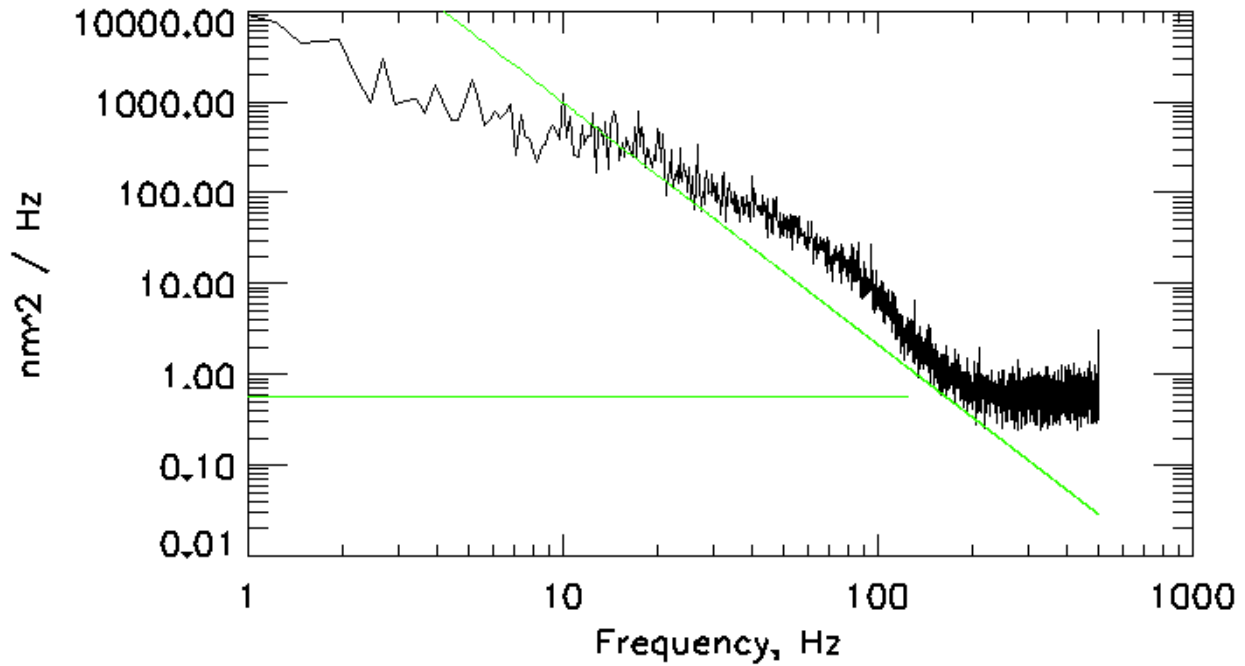

**Figure 12:** Example of bowed sky power spectrum. The black line is measurements from the open loop path, the horizontal line is the noise floor and the tilted line represents Kolmogorov theory. This type of shape was also seen in a third of the 40 data sets examined.

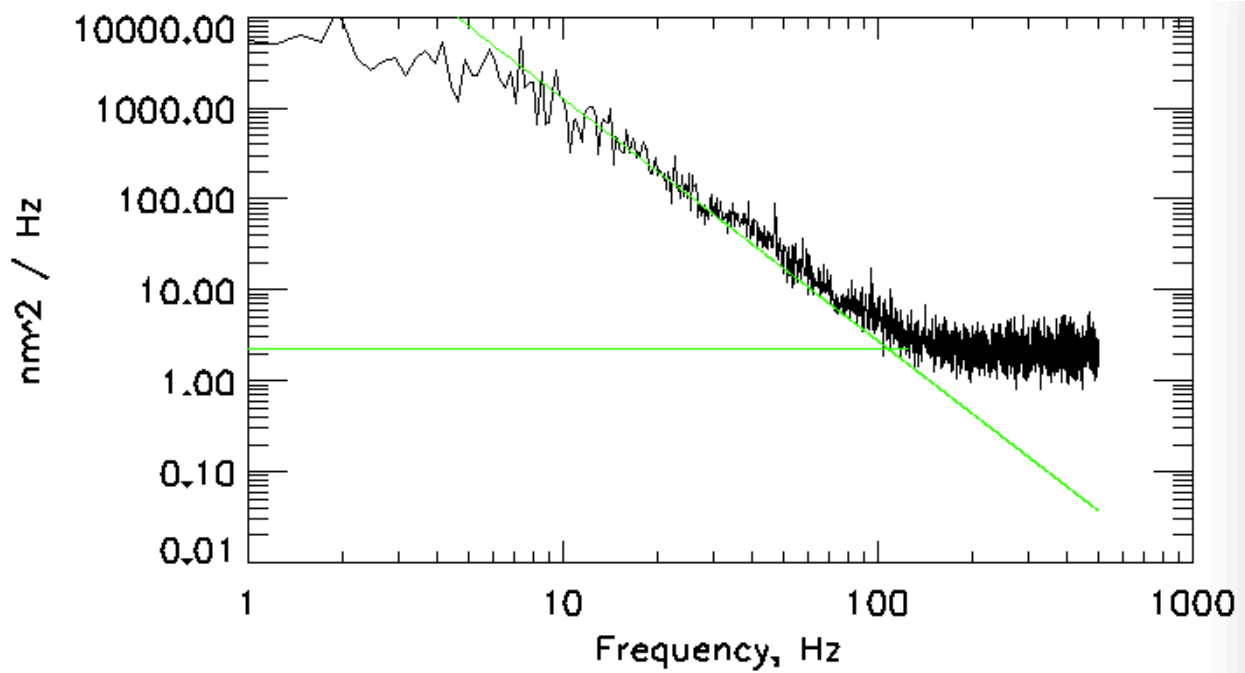

**Figure 13:** Example of sky power spectrum that follows Kolmogorov statistics well. The black line is measurements from the open loop path, the horizontal line is the noise floor and the tilted line represents Kolmogorov theory. This was only observed in a third of the 40 data sets examined.

The other main parameter of interest, $r_0$, was compared also. Based on 15 data sets from 6 nights between June and September, the average for the turbulence plate after being scaled to the 1 m diameter of the telescope was 9.9 cm.  The average for the sky, based on the 12 data sets when that exhibited Kolmogorov turbulence, was 8.7 cm.  These values assume a wavelength of 650 nm, a spot size of 1.6" and plate scale of 1.73"/pixel.  The plate scale was determined by translating a white light source across the pupil[16].  Based on QPI measurements of the turbulence plate, the expected $r_0$ of the plate should be 12.5 cm.  From previous measurements of the seeing at Mt. Hamilton determined by Don Gavel and Elinor Gates, the median $r_0$ of the sky is near 10 cm[17].  The deviation from expected of the actual measurements can be understood by considering the variability in parameters such as spot size, peak wavelength, and plate scale.  For example, decreasing the value of the plate scale by only 10% increases $r_0$ by ~13%.  Morzinski et al. found the spot size can range from a diffraction-limited 0.6" to 1.9" for the broad-band WFS CCD.  The close agreement between the $r_0$ for the plate and sky indicates that the plate successfully met its design requirements, which was to have an $r_0$ slightly larger than what is typical at this site.

## 6. Discussion

Although the processes we've developed have resulted in numerous successful plates with a variety of strengths, some issues remain a hindrance to fully expedited manufacturing.  As mentioned previously, achieving differing strengths of turbulence involves controlling certain variables.  Mainly, the distance between the plate and the can, the speed of the can, and the number of layers applied.  The amount of airflow present during drying has also shown to be important, and for best quality should be minimized.  Through repeat trials with the same parameters it has become apparent that other, currently uncontrollable, factors affect the outcome.  Most notably seems to be the fullness of the spray can, but it is possible such things as humidity and barometric pressure also come into play. Another very challenging obstacle has been dust particles getting trapped in the spray before it dries.  Maintaining a clean environment around and inside the spraying apparatus has proved beneficial, but moving it to a clean room would be the best solution.  This is viewed as an eventual future step in project development, along with upgrading the system to monitor and/or control the additional parameters mentioned.

Although certain aspects of the plate making process sporadically give rise to varying results, the relative simplicity, speed, and low cost of the materials, means that undergoing several trials to produce the desired plate is completely reasonable.  Applying a single coat of paint takes only a few minutes, and the time before a plate is ready for testing varies from several hours to several days, depending on the number of layers applied and the dry time between them.  In the event of excessive contamination or some other problem, the acrylic can be removed from the substrate very effectively with the strategic use of hot water, allowing it to be reused.

For not entirely well understood reasons, some locations on the handmade plates show a discrepancy between the power spectrum and structure function.  Thus far, two main types of peculiarities have been noticed.  In some cases the power spectrum closely agrees with Kolmogorov statistics and has a well-defined value of $r_0$, but the slope of the structure function is significantly less than the expected 5/3 making $r_0$ not clearly determinable.  The second oddity is when the structure function fits precisely but the power spectrum has a non-linear shape.  Such issues generally effect a small number of measurements on an otherwise good plate, and have normal looking reconstructed phases.  Clearly these occurrences signal a fluctuation of the statistics from Kolmogorov theory, perhaps similar to the real atmosphere. In spite of this occasional variability these plates are still valuable for scientific use.

## 7. Conclusion

Testing new astronomical adaptive optics systems requires the ability to simulate atmospheric turbulence in the laboratory. In this paper we have described an innovative method of producing fixed aberration phase plates that produce wavefront distortions with characteristics similar to atmospheric turbulence. We have also demonstrated a means of adjusting the statistical properties imprinted on the plates by varying the parameters involved in the application process. This method has achieved values of $r_0$ ranging from approximately 200 – 1200 μm measured at 632 nm (or 150 – 905 μm measured at 500 nm), and can also be used to produced plates with high order only and low order only turbulence. Both the substrates and spray are durable, lightweight materials, not highly sensitive to factors such as environment humidity or temperature, and are easy to keep clean. The acrylic also appears to have good longevity, evidenced by the turbulence characteristics of acrylic on glass plates created early on in this study remaining unchanged.

**Acknowledgements**

This work was funded by The Association of Universities for Research in Astronomy through the Gemini Planet Imager program. Special thanks to Jim Ward of Lick Observatory for building the spraying machine, to Brian Dupra and David Hilyard of Lick Observatory for measuring the transmission spectra, taking over the manufacturing of the plates, and making a diagram of the machine. Renate Kupke of the UCO/Lick Laboratory for Adaptive Optics has been a vital help with QPI.